\def\draftversion{false}

\RequirePackage{ifthen}
\ifthenelse{\equal{\draftversion}{true}}{
  \documentclass[prb,galley,showpacs,preprintnumbers,citeautoscript,
      amsmath,amssymb,longbibliography]{revtex4-2}
}{
  \documentclass[citeautoscript,floatfix,aps,prb,twocolumn,
      superscriptaddress,longbibliography]{revtex4-2}
}
\usepackage[utf8]{inputenc}
\usepackage{amsmath,latexsym,psfrag}
\usepackage[pdftex]{graphicx}
\usepackage{comment}
\usepackage[utf8]{inputenc}

\usepackage{ragged2e} 
\usepackage{subcaption} 

\usepackage{epstopdf}
\usepackage{natbib}
\usepackage{array}
\usepackage{amssymb}
\usepackage{dcolumn}
\usepackage{bm}

\usepackage{soul}  

\usepackage{float}
\usepackage{multirow}

\usepackage[usenames,dvipsnames]{color}

\soulregister\cite7
\usepackage[%
  colorlinks=true,
  urlcolor=blue,
  linkcolor=blue,
  citecolor=blue
]{hyperref}

\ifthenelse{\equal{\draftversion}{true}}{
  \usepackage{showlabels}
  \marginparwidth 2.7in
  \marginparsep 0.5in
  \newcounter{comm} 
  \def\commnext{\stepcounter{comm}}
  \def\commtext{{\bf\color{blue}[\arabic{comm}]}}
  \def\commmar{{\bf\color{blue}[\arabic{comm}]}}
  \def\msm#1{\commnext\marginpar{\small MS\commmar: #1}\commtext}

  \def\mlab#1{\marginpar{\small\bf #1}}
  
}{
  \def\dvm#1{}
  \def\cdm#1{}
  \def\msm#1{}
  \def\asm#1{}
  \def\miq#1{}
  \def\mlab#1{}
  
}

\begin{document}

\title{Strain dependence of the Bloch domain component in 180$^\circ$ domains in bulk PbTiO$_{3}$ from first-principles.}

\author{ Stephen Chege }
\email{schege882@gmail.com}

\affiliation{ Materials Modeling Group, 
	Department of Physics, Earth and Environmental Sciences,
	The Technical University of Kenya,
	52428-00200, Nairobi, Kenya.}  

\author{ Louis Bastogne }

\affiliation{ Theoretical Materials Physics, 
              Q-MAT, Universit\'e de Li\`ege, 
              All\'ee du 6 ao\^ut, 19, B-4000 Sart-Tilman, Belgium}

\author{ Fernando G\'omez-Ortiz }

\affiliation{ Theoretical Materials Physics, 
              Q-MAT, Universit\'e de Li\`ege, 
              All\'ee du 6 ao\^ut, 19, B-4000 Sart-Tilman, Belgium} 

\author{ James Sifuna}
\affiliation{ Materials Modeling Group,
Department of Physics, Earth and Environmental Sciences,
The Technical University of Kenya, 52428-00200, Nairobi, Kenya.} 
\affiliation{ Theoretical Condensed Matter Group, Department of Natural Science,  The Catholic University of Eastern Africa, 62157 - 00200, Nairobi, Kenya.}

\author{ George Amolo }
\affiliation{ Materials Modeling Group, 
	Department of Physics, Earth and Environmental Sciences,
	The Technical University of Kenya,
	52428-00200, Nairobi, Kenya.}  

\author{ Philippe Ghosez }

\affiliation{ Theoretical Materials Physics, 
              Q-MAT, Universit\'e de Li\`ege, 
              All\'ee du 6 ao\^ut, 19, B-4000 Sart-Tilman, Belgium}

\author{ Javier Junquera }
\email{javier.junquera@unican.es}
\affiliation{ Departamento de Ciencias de la Tierra y
	F\'{\i}sica de la Materia Condensada, Universidad de Cantabria,
	Avenida de los Castros s/n, 39005 Santander, Spain.}

\date{\today}

\begin{abstract}
We investigate the emergence of Bloch-type polarization components in 180$^\circ$ ferroelectric domain walls in bulk PbTiO$_{3}$ under varying mechanical boundary conditions, using first-principles simulations based on density functional theory. A spontaneous Bloch component—primarily associated with Pb displacements confined within the PbO domain wall plane—can condense under realistic strain conditions on top of the Ising-type domain walls.  The amplitude and energetic stabilization of this component are highly sensitive to the in-plane lattice parameters. In particular, tensile strains akin to those imposed by DyScO$_{3}$ substrates enhance the Bloch component and lead to energy reductions as large as 10.7 mJ/m$^{2}$ (10.6 meV/$\square$) with respect to the most stable structure including only Ising and N\'eel components. We identify a relatively flat energy landscape for the Bloch polarization,  highlighting the tunability of chiral textures through strain engineering. Our results offer a predictive framework for estimating the strain-dependent onset temperature of Bloch-type domain wall components and provide insight into the design of topologically nontrivial and chiral polar structures in ferroelectrics.
\end{abstract}

\maketitle
\section{Introduction}
\label{sec:introduction}

For a long time, it was widely believed that ferroelectrics could not sustain intricate topological arrangements of electric dipoles.
The strong coupling between polarization and the lattice was thought to impose significant structural and dipolar anisotropy, making polarization rotation prohibitively expensive in terms of energy.

However, pioneering theoretical studies on nanodots and nanodisks in the early 2000s~\cite{Fu-03,Naumov-04}, followed by combined theoretical and experimental investigations, have challenged this view. These efforts uncovered a variety of nontrivial topological structures—such as flux-closure patterns~\cite{Jia-11,Tang-15}, vortices~\cite{Yadav-16}, electric skyrmion bubbles~\cite{Nahas-15,Pereira-19,Das-19}, merons~\cite{Nahas-20.2,Wang-20,Shao-23}, and hopfions~\cite{Lukyanchuk-20}, among others—laying the foundation for an emerging field.

In low-dimensional ferroelectric oxide nanostructures, the interplay between electrostatic energies (arising from confinement and depolarization effects), elastic contributions, and gradient terms can significantly influence polarization patterns, leading to the stabilization of topologically nontrivial polar structures.
For a more comprehensive overview of this rapidly evolving research area, readers may refer to Ref.~\cite{Junquera-23,Govinden-23,Wang-23}.

One of the most extensively studied systems in this context is the PbTiO$_{3}$/SrTiO$_{3}$ heterostructure.
As a consequence of the complex polar orderings, exotic functional properties may emerge as it is the case of the negative capacitance effect~\cite{Zubko-16,Yadav-19,Iniguez-19}.
Another interesting example is related with the appearance of chirality~\cite{Shafer-18,Lovesey-18,Kim-22,McCarter-22}, even though both constituent materials are achiral in their bulk forms.
The intricate relationship between chirality and topology in ferroelectric nanostructures has been extensively reviewed in Ref.~\cite{Lukyanchuk-24,Lukyanchuk-25}.

Chirality, which describes a structural asymmetry where an object cannot be superimposed onto its mirror image, is a fundamental property with profound implications across multiple disciplines, including chemistry, physics, mathematics, and biology~\cite{Felser-22,Bousquet_2025}.
In either magnetically or electrically polar systems, the presence of a vortex with a nonzero toroidal moment does not necessarily imply chirality in three dimensions. Depending on the symmetry of the mirror plane, a vortex structure may be either superimposable onto its mirror image or require only a rigid translation of half a unit cell. As a result, such vortices exhibit chirality primarily in two-dimensional systems, where symmetry transformations are constrained to those preserving dipole orientations within the plane.

In magnetism, long-range chiral ordering typically originates from the Dzyaloshinskii-Moriya interaction (DMI)~\cite{Dzyaloshinsky-58,Moriya-60}.
Only recently have electric analogs of this interaction been proposed~\cite{Erb-20,Zhao-21,Chen-22}, where oxygen octahedral tilting mediates the coupling of an electric DMI, playing a role analogous to spin-orbit coupling in magnetic systems.
However, this mechanism is absent in both bulk PbTiO$_{3}$ and PbTiO$_{3}$/SrTiO$_{3}$ heterostructures.

Crucially, ferroelectric systems do not require DMI-like interactions to exhibit chirality.
For example, chiral behavior in electric polar vortices can arise from the coupling between an axial polarization component (perpendicular to the vortex plane) and the vorticity of clockwise or counterclockwise vortices.
This phenomenon was first predicted in BaTiO$_{3}$/SrTiO$_{3}$ nanocomposites using first-principles-based effective Hamiltonians~\cite{Louis-12}.
The emergence of this axial polarization component is linked to the condensation of Bloch domain walls in ferroelectric domains~\cite{Wojdel-14}.
Specifically, a 180$^\circ$ ferroelectric domain wall develops a Bloch component when the polarization rotates within a plane parallel to the domain wall~\cite{Lee2009}.
Given its fundamental implications, several studies have explored the formation and stability of Bloch-type domain walls.

Surprisingly, early first-principles calculations on 180$^\circ$ domain walls in bulk PbTiO$_{3}$ did not detect the presence of a Bloch polarization component~\cite{MeyerVanderbilt2002}.
The reasons remain unclear but may stem from symmetry constraints imposed during structural relaxations to reduce computational cost or from the choice of initial geometry, which could trap the system in a local energy maximum.
The first {\it ab-initio} evidence of a switchable Bloch component in PbTiO$_{3}$ domain walls was reported by Wojde\l{} and \'I\~niguez~\cite{Wojdel-14}, using first-principles calculations at $T=0$ K with a PBEsol-based functional~\cite{Perdew-08}.
They observed a Bloch polarization component of approximately 0.4 $\rm C/m^{2}$ at the domain wall plane, which rapidly vanished within the adjacent domains.
Its origin was attributed to the large Pb atomic displacements and Pb-O hybridization at the domain wall~\cite{Wojdel-14,Wang-14,Wang-17}.

The development of Bloch-like domain walls plays a crucial role in the condensation of polar Bloch skyrmions in single-phase PbTiO$_{3}$~\cite{Pereira-19}.
It is also linked to the emergence of vortex tubes in SrTiO$_{3}$/PbTiO$_{3}$ heterostructures grown on DyScO$_{3}$ substrates, endowing them with a distinct handedness~\cite{Shafer-18}.
Notably, these chiral properties can be externally controlled via an applied electric field~\cite{Behera-22}.
The condensation of the Bloch component in PbTiO$_{3}$/SrTiO$_{3}$ heterostructures at sufficiently low temperatures has been confirmed using first-principles~\cite{Aguado-Puente-12}, second-principles~\cite{Shafer-18}, deep potential molecular dynamics~\cite{Yang-24}, and phase-field perturbation models~\cite{Zheng-25}.

However, recent studies suggest that the predicted Bloch polarization component in 180$^\circ$ domain walls within PbTiO$_{3}$/SrTiO$_{3}$ heterostructures may be more sensitive to boundary conditions than previously thought~\cite{Zatterin-24}.
Key factors such as epitaxial strain from the substrate, layer thickness, temperature, and equilibrium domain periodicity significantly influence its stabilization and can even suppress its formation.
This complexity is further highlighted by the observation of N\'eel-type domain structures in PbTiO$_{3}$ thin films at room temperature, where a clear Bloch component is absent, suggesting at best a metastable nature~\cite{Weymann-22}.

Among these factors, epitaxial strain is particularly influential due to the strong polarization-strain coupling.
The influence of strain and the reduced dimensionality on the energetics of the domain-wall confined Bloch ferroelectric distortion was already emphasized in Ref.~\cite{Wojdel-14}.
Also, some studies have demonstrated that the phase diagram of PbTiO$_{3}$/SrTiO$_{3}$ superlattices, under short-circuit electrical boundary conditions, evolves as a function of substrate lattice constant.
The system transitions from a tetragonal phase under large compressive strain to polar skyrmion bubbles (with SrTiO$_{3}$-imposed strain), then to vortex phases (with DyScO$_{3}$-imposed strain), and finally to $a_{1}/a_{2}$ domain structures under large tensile strain~\cite{Junquera-23}.
Even coexistence of $a_{1}/a_{2}$ and vortex phases has been reported on DyScO$_{3}$ substrates~\cite{Damodaran-17.2}, while single-phase chiral vortex structures can be achieved by controlling epitaxial strain~\cite{Das-23}.

In this work, we explore the influence of mechanical boundary conditions on the stability of the Bloch component in 180$^\circ$ domain walls in bulk PbTiO$_{3}$ using first-principles simulations.

\section{Methodology}
\label{sec:methodology}

The simulations of 180$^\circ$ domains in bulk PbTiO$_{3}$ have been carried out with {\sc siesta} within the generalized gradient approximation to the density functional theory, using the PBEsol functional~\cite{Perdew-08}. 
Core electrons were replaced by {\it ab initio} norm-conserving fully separable pseudopotentials~\cite{Kleinman-82}.
In this work the optimized norm-conserving Vanderbilt pseudopotentials proposed by Hamann~\cite{norm_conserving} were used,
in the {\sc psml} format~\cite{psml} available in the Pseudo-Dojo periodic table~\cite{pseudodojo,footnotepseudo}.

The one-electron Kohn-Sham eigenstates were expanded in a basis of strictly localized~\cite{Sankey-89} numerical atomic orbitals~\cite{Artacho-99}.
Basis functions were obtained by finding the eigenfunctions
of the isolated atoms confined within the soft-confinement spherical potential proposed in Ref.~\cite{Junquera-01}.
A quadruple-$\zeta$, double polarized basis set was used for the valence states of all the atoms. 
For the semicore states of 
Pb (5$s$, 5$p$, and 5$d$), and Ti (3$s$, and 3$p$), a double-$\zeta$ quality was chosen.
For Pb, an extra single-$\zeta$ 5$f$ shell was included to increase angular flexibility.
All the parameters that define the range and the shape of the atomic orbitals have been optimized variationally in bulk cubic centrosymmetric PbTiO$_{3}$ following the recipe given in Ref.~\cite{Junquera-01}.

The electronic density, Hartree, and exchange-correlation potentials, as well as the corresponding matrix elements between the basis orbitals, were calculated in a uniform real space grid~\cite{Soler-02}. An equivalent plane-wave cutoff of 1200 Ry was used to represent the charge density.

To assess the convergence of the numerical atomic orbital (NAO) basis set, we compared the results obtained with {\sc siesta} against those from a highly converged plane-wave calculation performed with {\sc abinit}~\cite{gonze2009abinit,gonze2016,gonze2020}, using the same exchange-correlation functional and pseudopotentials. The plane-wave results, with a very high energy cutoff (65 Ha), serve as a reference for the complete basis-set limit. Table-\ref{table:structtetra-fin} presents the lattice constants, atomic displacements, and spontaneous polarization of fully relaxed bulk tetragonal PbTiO$_3$ in its ferroelectric phase.
Overall, we observe good agreement between the two codes and approaches. The tetragonality is slightly underestimated in {\sc siesta} by approximately $-0.8$\%. The largest discrepancy in atomic displacements occurs for the apical oxygen atom O$_{\mathrm{III}}$, where the displacement along the $z$ direction [$\xi_{z}$(O$_{\rm III}$)] is underestimated by $-5.9$\%.
The bulk spontaneous polarization can be computed as the change of polarization, along a continuous distortion path -- from the centrosymmetric non-polar phase to the polar phase -- using the Berry phase approach~\cite{Vanderbilt-book}. This formal description as defined by the Berry-phase theory, yields a value of $0.96$~C/m$^{2}$ with {\sc siesta}, in comparison to $0.99$~C/m$^{2}$ from {\sc abinit}.  Assuming constant Born effective charges along this path, the polarization change can alternatively be approximated by the sum of the product of the Born effective charges times the associated atomic displacements, divided by the unit cell volume~\cite{Ghosez-98}.
Using Born effective charges calculated in the centrosymmetric cubic phase (resp. tetragonal phase), we obtain an effective polarization of $1.14$~C/m$^{2}$ (resp. 0.84~C/m$^{2}$) from both codes. The overestimation (resp. underestimation) relative to the Berry-phase polarization arises from the sensitivity of the Born charges to atomic displacements. More accurate estimates could be obtained by averaging the Born effective charges evaluated in both the high-symmetry and distorted structures.
These differences will have implications for the layer-resolved polarization discussed in Sec.~\ref{sec:dwstructure}.

The five atom unit cell was replicated 20 times along the [100] direction, building a supercell consisting of $20\times 1 \times 1$ perovskite unit cell (100 atoms in the simulation box).  Periodic boundary conditions along the three directions of space were applied in the supercell. 
This corresponds to performing the simulation under short-circuit electrostatic conditions. 
The domain wall is centered on a PbO plane, which is the energetically preferred configuration, as it exhibits a lower domain wall energy than the TiO$_2$-centered alternative~\cite{MeyerVanderbilt2002}.
Then, the bulk soft mode distortion shown in Table-\ref{table:structtetra-fin} was superimposed
on the PbTiO$_{3}$ atoms, so the polarization points upwards in half
of the superlattice and downward in the other half.
We take the domain wall to lie in the $yz$ plane.
With this choice of the coordinate system, the polarization within the interior of a domain points along the $z$ direction, while the domain-wall Bloch polarization lies along the $y$ direction and the N\'eel component along the $x$ direction.

A $1 \times 8 \times 8$ $\bf{k}$-point Monkhorst-Pack grid~\cite{Monkhorst-76} was used for integrations within the first Brillouin zone of the supercell, and $8 \times 8 \times 8$ for the simulations of the monodomain phases with five atoms per unit cell. 

In order to study the stability of the different domains configurations, we shall pay attention to three different magnitudes.
The first is the energy cost required to form an Ising domain without any N\'eel or Bloch component with respect to the homogeneous ferroelectric phase. This will be referred as the Ising domain wall energy, $E_{\rm DW}^{\rm Ising}$. It is computed from

\begin{equation}
   E_{\rm DW}^{\rm Ising} = \frac{1}{2A}
     \left(E_{\rm sc}^{\rm Ising} - 20 \times E_{\rm bulk}^{\rm ferro} \right),
     \label{eq:DW}
\end{equation}

\noindent where $E_{\rm sc}^{\rm Ising}$ is the energy of the relaxed supercell where only the Ising domain is allowed to condense (i.e. atomic relaxation along $z$  only), $E_{\rm bulk}^{\rm ferro}$ is the reference energy of the bulk structure (without domain walls),  calculated from a single perovskite unit cell in the relaxed ferroelectric phase using a $k$-point mesh equivalent to the one used for the supercell.
The factor of 1/2 is due to the fact that in a supercell with periodic boundary conditions there are two domain walls. $A$ represents the area of the domain wall, computed from the corresponding lattice parameter within the supercell. 
Since the most stable structure in bulk PbTiO$_{3}$ under short-circuit boundary conditions is the monodomain ferroelectric phase, $E_{\rm DW}^{\rm Ising}$ is positive.
Domain walls energies will be given in units of mJ/m$^{2}$ or in units of meV/$\square$, where $\square$ represents the cell surface area of the domain wall. 
 
The second relevant magnitude is the energy gain for condensing \emph{only} the N\'eel component at the domain wall on top of the Ising configuration (i.e. atomic relaxation along $z$ and $x$). 
This is referred from now on as $E_{\rm DW}^{\rm N\acute{e}el}$.
It is computed applying an equivalent relation as the one described by Eq.~(\ref{eq:DW}) with $E_{\rm sc}^{\rm Ising}$ replaced by $E_{\rm sc}^{\rm N\acute{e}el}$, the energy of the relaxed structure allowing for a N\'eel component  at the domain wall. 
During this process, the appearance of a Bloch component of the polarization at the domain wall is constrained.
%
%

Finally, if on top of the most stable Ising and N\'eel components of the polarization at the domain wall, we further allow the condensation of the Bloch component (i.e. atomic relaxation along $x$, $y$ and $z$), we can define $E_{\rm DW}^{\rm Bloch}$, again applying an equivalent relation as the one described by Eq.~(\ref{eq:DW}), but introducing $E_{\rm sc}^{\rm Bloch}$ as the energy of the most stable ferroelectric polarization (including both N\'eel and Bloch components).
The difference between  $E_{\rm DW}^{\rm Bloch}$ and $E_{\rm DW}^{\rm N\acute{e}el}$ is particularly relevant to estimate the critical temperature upon which the Bloch component is unstable $T_{\rm C}^{\rm DW}$.
Structural relaxations allowing for Bloch components were initialized from the previously relaxed N\'eel configuration, with Pb atoms at the domain wall displaced along the $y$-direction. Subsequent relaxations were performed using a conjugate-gradient method. Unless otherwise specified, all atomic relaxations were carried out using the {\sc siesta} code, with convergence criteria of 
0.01 eV/\AA\ for the maximum component of the force on any atom and 0.0001 eV/\AA$^{3}$ for the maximum component of the stress tensor. For the atomic relaxation carried out using the {\sc abinit} code, a convergence criteria of 2.5$\times$10$^{-3}$~eV/\AA\ for the maximum component of the force on any atom and 2.5$\times$10$^{-5}$ eV/\AA$^{3}$ for the maximum component of the stress tensor has been used.

\begin{table}[!h]
   \begin{center}
   \caption{\justifying
       Lattice constants ($a$ and $c$) and atomic displacements of bulk PbTiO$_3$ in its ferroelectric tetragonal phase. The atomic displacements, measured relative to the centrosymmetric tetragonal positions, are expressed as a unit vector $\hat{\xi}$ multiplied by an amplitude $\vert \vec{\xi} \vert$, following the approach in Ref.~\cite{MeyerVanderbilt2002}. $\Delta E$ represents the energy difference between the relaxed ferroelectric tetragonal structure and the centrosymmetric cubic phase. $P_{z}^{\rm BP}$ denotes the polarization magnitude along the $z$-axis in the ferroelectric tetragonal unit cell computed from the Berry phase, while $P_{z}^{\rm BEC}$ is the polarization computed with the product of the Born effective charges computed at the cubic phase times the displacements of the atoms. Experimental data are sourced from Ref.~\cite{Jona-93}.
      }
   \label{table:structtetra-fin}
       \begin {tabular}{lccc}
       \hline
       \hline
                                 &
       NAO                       &
       PW                        &
       Expt.                     \\
       \hline
       $a$ (\AA)                 &
       3.871                     &
       3.861                     &
       3.905                     \\
       $c$ (\AA)                 &
       4.220                     &
       4.244                     &
       4.151                     \\
       $c/a$                     &
       1.090                     &
       1.099                     &
       1.063                     \\  
       $\xi_{z}$ (Pb)            &
       0.696                     &
       0.699                     &
       0.718                     \\
       $\xi_{z}$ (Ti)            &
       0.368                     &
       0.365                     &
       0.335                     \\
       $\xi_{z}$ (O$_{\rm I}$)   &
       -0.373                    &
       -0.363                    &
       -0.351                    \\  
       $\xi_{z}$ (O$_{\rm II}$)  &
       -0.373                    &
       -0.363                    &
       -0.351                    \\  
       $\xi_{z}$ (O$_{\rm III}$) &
       -0.318                    &
       -0.338                    &
       -0.351                    \\
       $\sum_{i} \xi_{i}$        &
       0.000                     &
       0.000                     &
       0.000                     \\
       $\vert \xi \vert$ (\AA)   &
       0.490                     &
       0.510                     &
       0.434                     \\ 
       $\Delta E$ (meV)          &
       -79.38                    &
       -85.24                    &
                                 \\ 
       $P_{z}^{\rm BEC}$ (C/m$^{2}$) &
       1.14                      &
       1.14                      &
                                 \\
       $P_{z}^{\rm BP}$ (C/m$^{2}$)  &
       0.96                      &
       0.99                      &
                                 \\ 
       \hline
       \hline
       \end {tabular}
   \end{center}
\end{table}

\section{Results}
\label{sec:results}

\subsection{Atomistic domain-wall structures}
\label{sec:dwstructure}

The relaxed atomic structures of the supercell were obtained under various epitaxial strain conditions. These include: (i) full relaxation of both atomic positions and lattice vectors; (ii) relaxation with lattice constants fixed to those of the optimized bulk tetragonal ferroelectric phase (see Table-\ref{table:structtetra-fin}); (iii) relaxation under the epitaxial constraint imposed by a hypothetical SrTiO$_3$ substrate; and (iv) relaxation under strain conditions corresponding to a DyScO$_3$ substrate, both assuming a square surface unit cell, or an orthorhombic one with strains of -0.25 \% and -0.16 \% with respect the theoretical bulk cubic lattice constant along the two in-plane directions~\cite{Tovaglieri-25}. The resulting lattice parameters for each case are summarized in Table-\ref{table:DW-energy}.

\begin{figure*}[htbp]
    \centering
    \includegraphics[width=0.95\columnwidth]{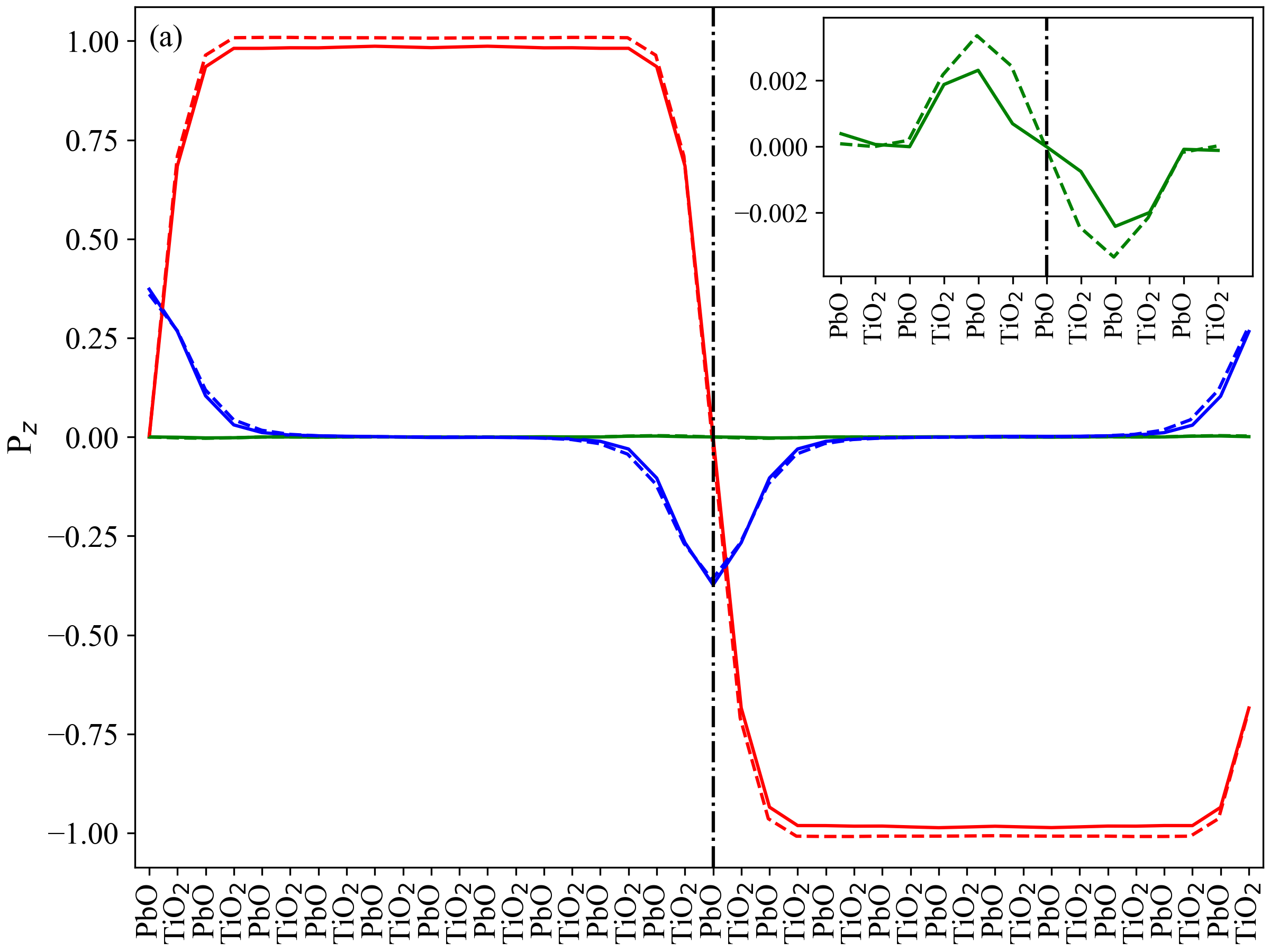}
    \hspace{0.5cm}
    \includegraphics[width=0.95\columnwidth]{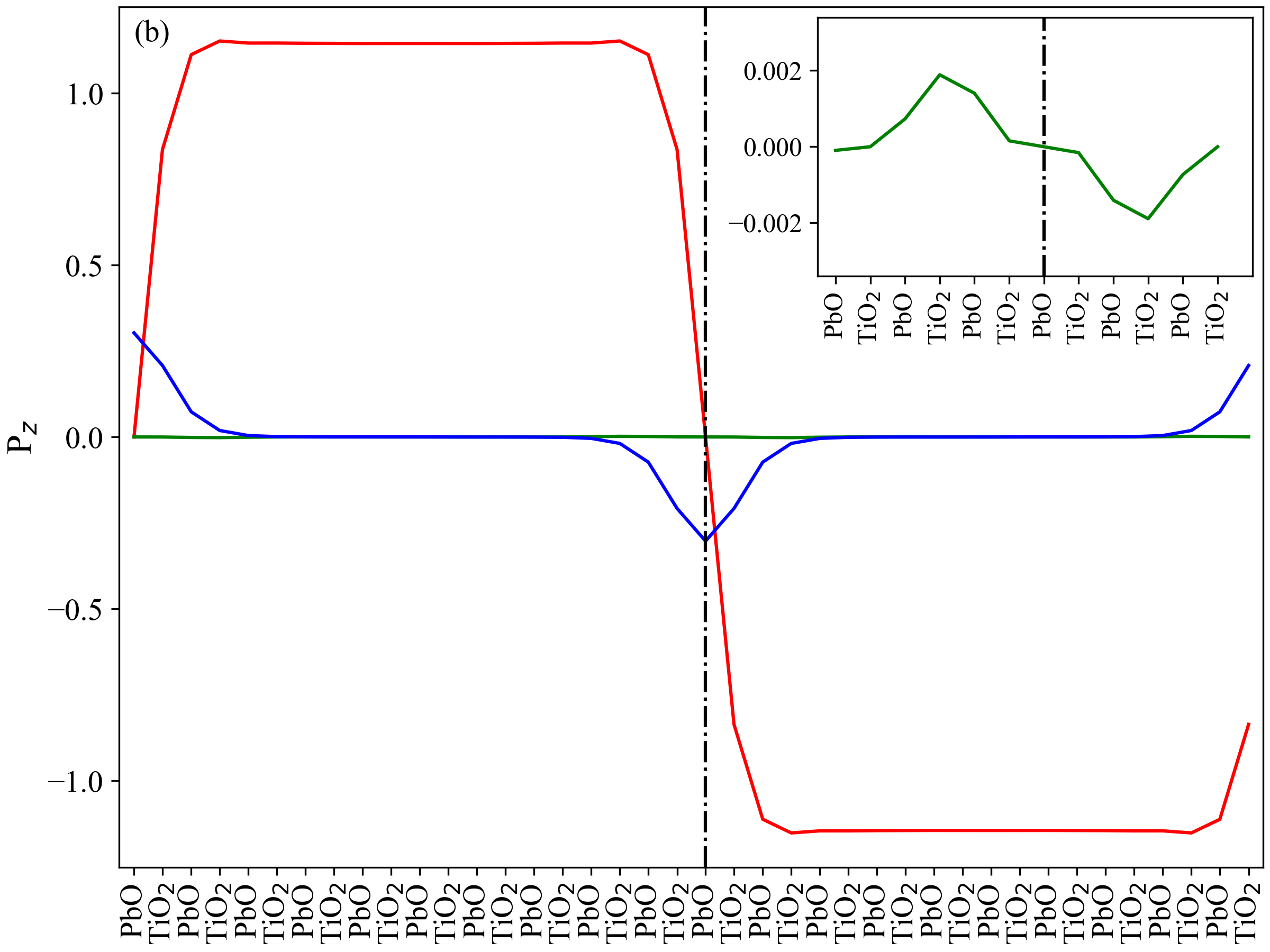}

    \vspace{0.5cm}
    \includegraphics[width=0.95\columnwidth]{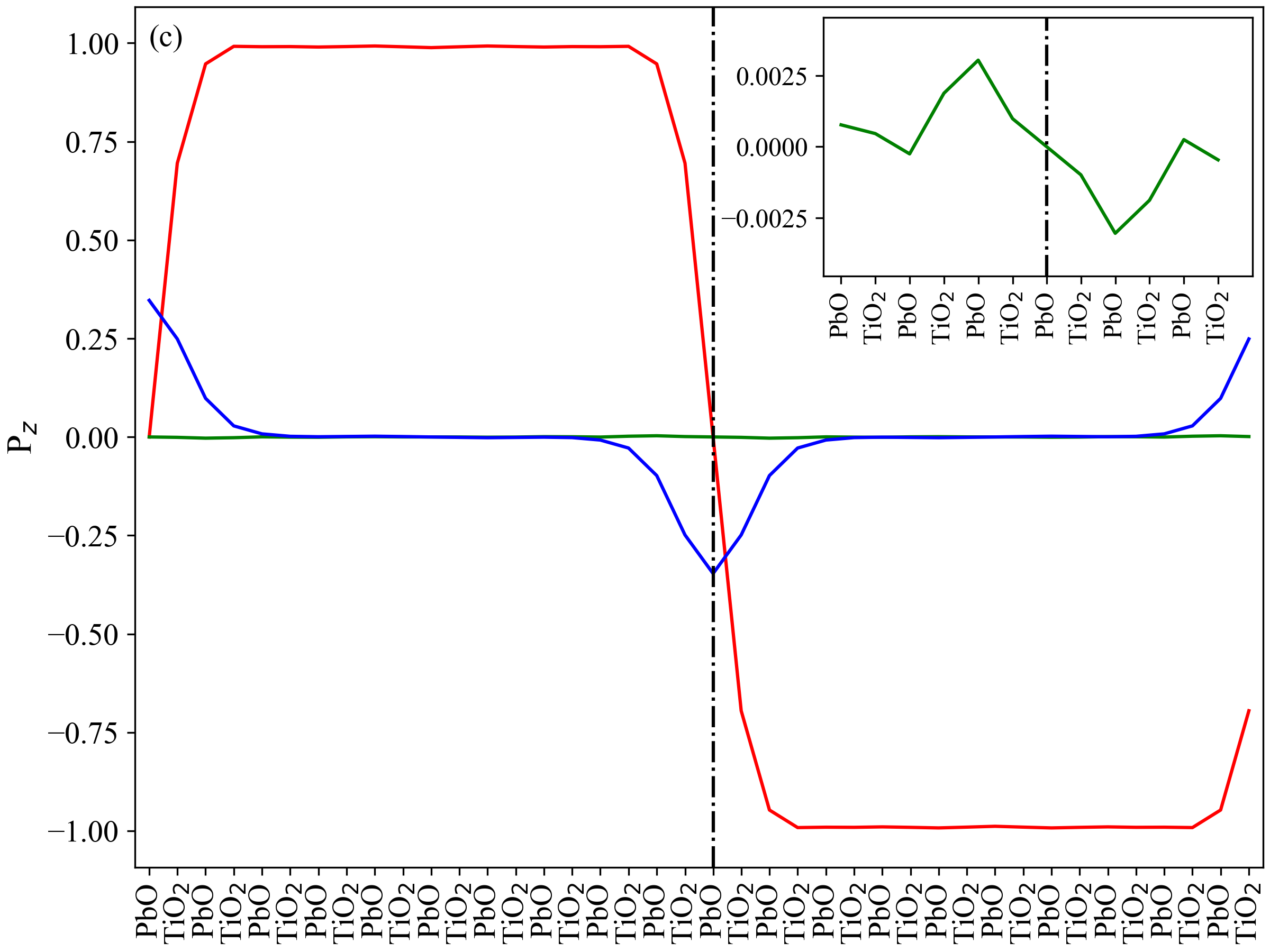}
     \hspace{0.5cm}
     \includegraphics[width=0.95\columnwidth]{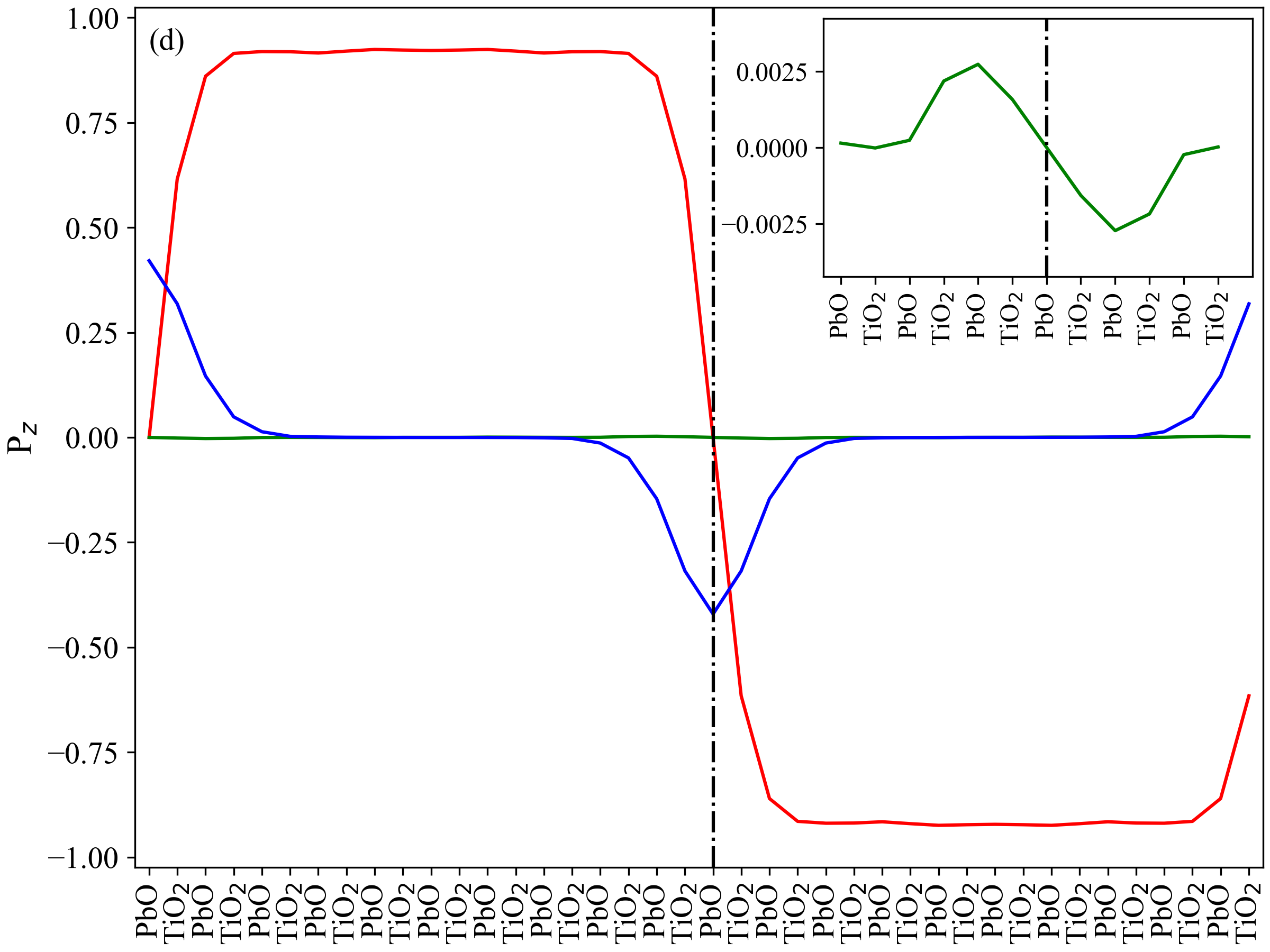}

     \vspace{0.5cm}
     \includegraphics[width=0.95\columnwidth]{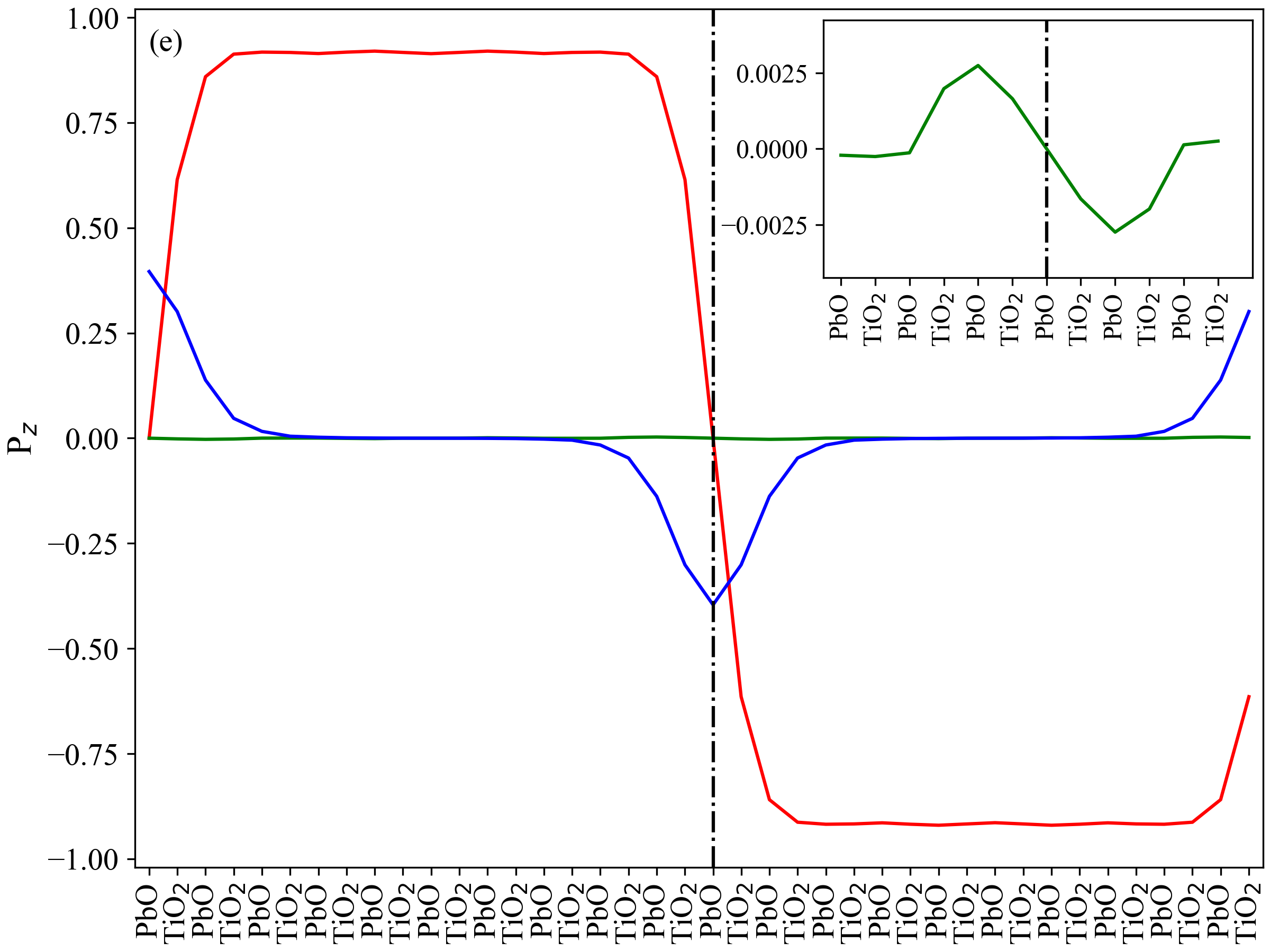}
     \hspace{0.5cm}
     \includegraphics[width=0.95\columnwidth]{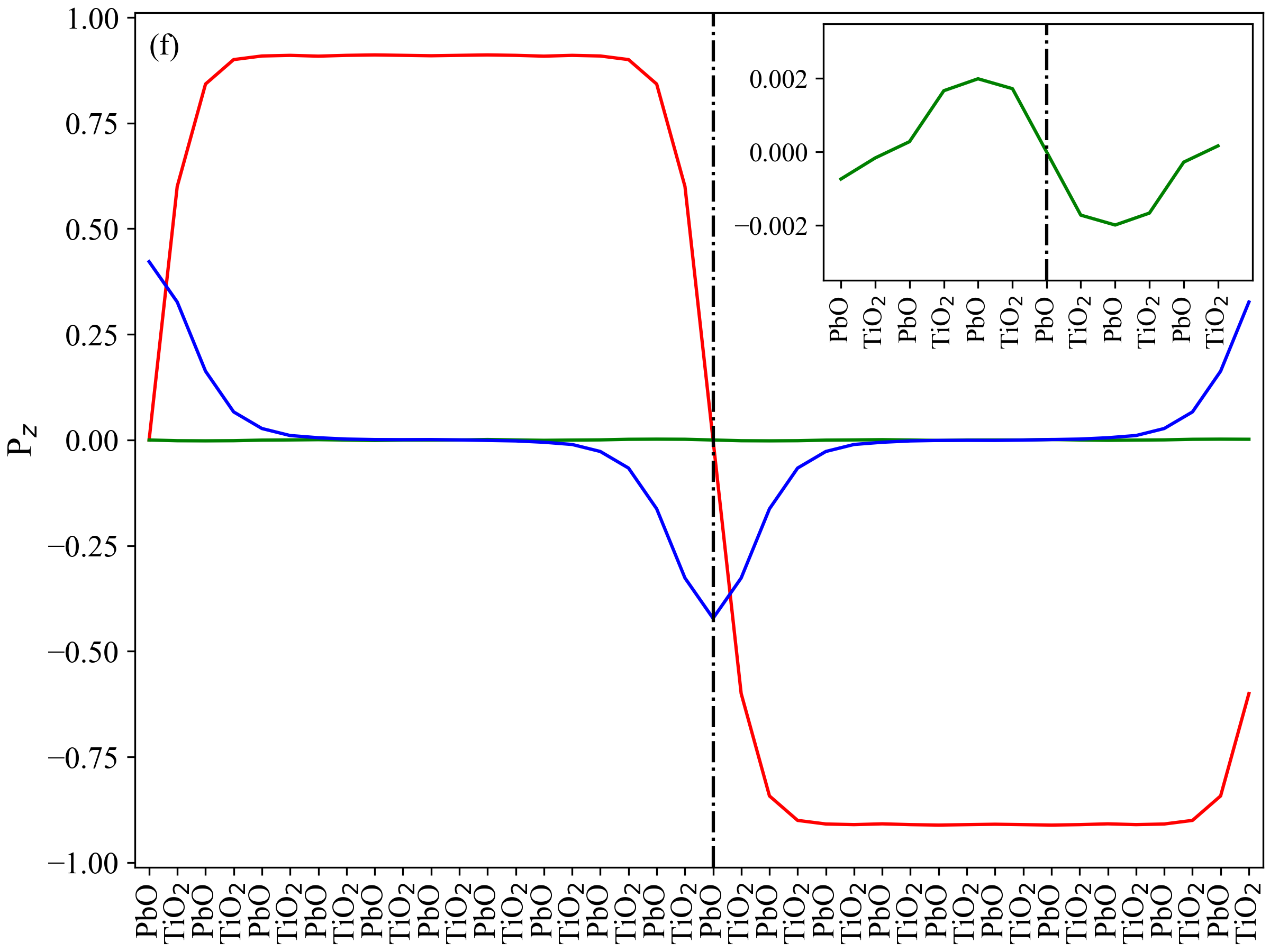}
    \caption{\justifying Layer-resolved polarization profiles across 180$^\circ$ ferroelectric domains in bulk PbTiO$_3$ under six different epitaxial strain conditions: (a) full relaxation of atomic positions and lattice vectors; (b) relaxation with lattice parameters fixed to those of the optimized bulk tetragonal ferroelectric phase (see Table-\ref{table:structtetra-fin}); (c) relaxation under an epitaxial constraint mimicking a hypothetical SrTiO$_3$ substrate ($a = b = 3.889$ \AA) ; (d) relaxation under strain corresponding to a DyScO$_3$ substrate with a square surface unit cell ($a = b = 3.912$ \AA); (e) relaxation with orthorhombic DyScO$_3$ substrate constraints, $a = 3.912$ \AA\ and $b = 3.916$ \AA; and (f) the same as (e) but with $a = 3.916$ \AA\ and $b = 3.912$ \AA.
    Components of the polarization along the three cartesian directions are represented by green ($P_{x}$), blue ($P_{y}$) and red ($P_{z}$) lines. The full (respectively dashed) lines correspond to the results obtained with {\sc siesta} (respectively {\sc abinit}~\cite{Zatterin-24}). Vertical dot-dashed lines mark the position of the PbO plane at the domain wall. 
    Inset: zoom on the $P_{x}$ component around the domain wall. Polarizations are given in C/m$^{2}$.}
    \label{fig:structure}
\end{figure*}

The corresponding layer-resolved polarization profiles are presented in Fig.~\ref{fig:structure}.
The local polarization is obtained within a linear approximation, using the same approach described in Sec.~\ref{sec:methodology} for the computation of the spontaneous polarization~\cite{MeyerVanderbilt2002,Wojdel-14} in usual first-principles models of ferroelectrics~\cite{Zhong-95}. The polarization is computed, as a product of the Born effective charges obtained for the bulk cubic centrosymmetric phase of PbTiO$_{3}$ times the displacement of the atoms with respect to a reference centrosymmetric phase on a five atom unit cell centered on Ti or Pb depending whether we are on a PbO or a TiO$_2$ planes.

Several conclusions can be drawn from the data in Fig.\ref{fig:structure}. First, the comparison between results obtained using {\sc siesta} and {\sc abinit} for the fully relaxed structure [Fig.~\ref{fig:structure}(a)] confirms excellent agreement for all three components of the polarization, validating the use of numerical atomic orbitals in domain wall calculations.

Second, consistent with prior findings~\cite{Catalan-12}, the 180$^\circ$ ferroelectric domain walls are atomically narrow, with a width on the order of the lattice constant. 
The domain wall width was extracted by fitting the polarization profile to the functional form proposed by Gureev {\it et al.}~\cite{Gureev-11},

\begin{equation} 
   P(x) = P_{0} \tanh \left( \frac{x}{\delta} \right), 
   \label{eq:funcform} 
\end{equation}

\noindent where $\delta$ denotes the characteristic half-width of the domain wall (i.e., to make the transition from $-P_{0}$ to $P_{0}$ we need a space of around four times this length $\delta$). The results, reported in Table~\ref{table:DW-energy}, reveal that $\delta$ increases with the in-plane lattice parameter, ranging from 2.06~\AA\ at $a = 3.871$~\AA\ to 2.45~\AA\ at $a = 3.916$~\AA.

Third, the layer-by-layer polarization exhibits a plateau beginning at the second TiO$_{2}$ plane, with a saturation value of $P_{0} = 0.99$ C/m$^2$ for the fully relaxed structure. This is somewhat lower than the bulk value of 1.14 C/m$^2$ reported in Table-\ref{table:structtetra-fin}, calculated using the same local polarization model. This discrepancy arises from  the compromise that should be found between the $c$ value at the domain and the $c$ value at the domain wall. The $c$ value obtained after relaxation of the full structure is $c = 4.115$~\AA, i.e., 2.49\% shorter than the relaxed value for the bulk tetragonal phase and is dependent of the width of the domains. The reason for this shortening can be traced back to the strong polarization-strain coupling: the local $c$ would naturally reduce near the domain wall where $P_{z}$ vanishes and would tend to the bulk value at the center of the domain, where $P_{z}$ saturates. 
Thus, a balance is found where the global $c$ is reduced respective to the bulk, with a concomitant increase of the in-plane lattice constant 
($a$ changes from 3.871 \AA\ in the bulk ferroelectric tetragonal phase to 3.892 \AA\ in the fully-relaxed domain structure). 
Indeed, the local in-plane strain component, $\varepsilon_1$ in Voigt notation, computed with respect to the bulk tetragonal ferroelectric in-plane lattice constant, shows a uniform tensile strain of approximately 0.54\% within the domain interior. This effect diminishes with increasing simulation box length along $x$, so the contribution to the total energy coming from the center of the domain increases with respect to the regions close to the domain wall. 

We notice that constraining the lattice constants to that of the bulk tetragonal phase, as done in Ref.~\cite{MeyerVanderbilt2002}, restores   a central domain polarization value of $P_{0} = 1.15$~C/m$^2$, in excellent agreement with the bulk value [Fig.~\ref{fig:structure}(b)], independently of the domain width.
Notably, $P_{0}$ shows also a strong dependence on epitaxial strain: compressive strain (smaller in-plane lattice constant) enhances the polarization amplitude at the center of the domains.

\begin{figure}[H]
    \centering
    \includegraphics[width=\columnwidth]{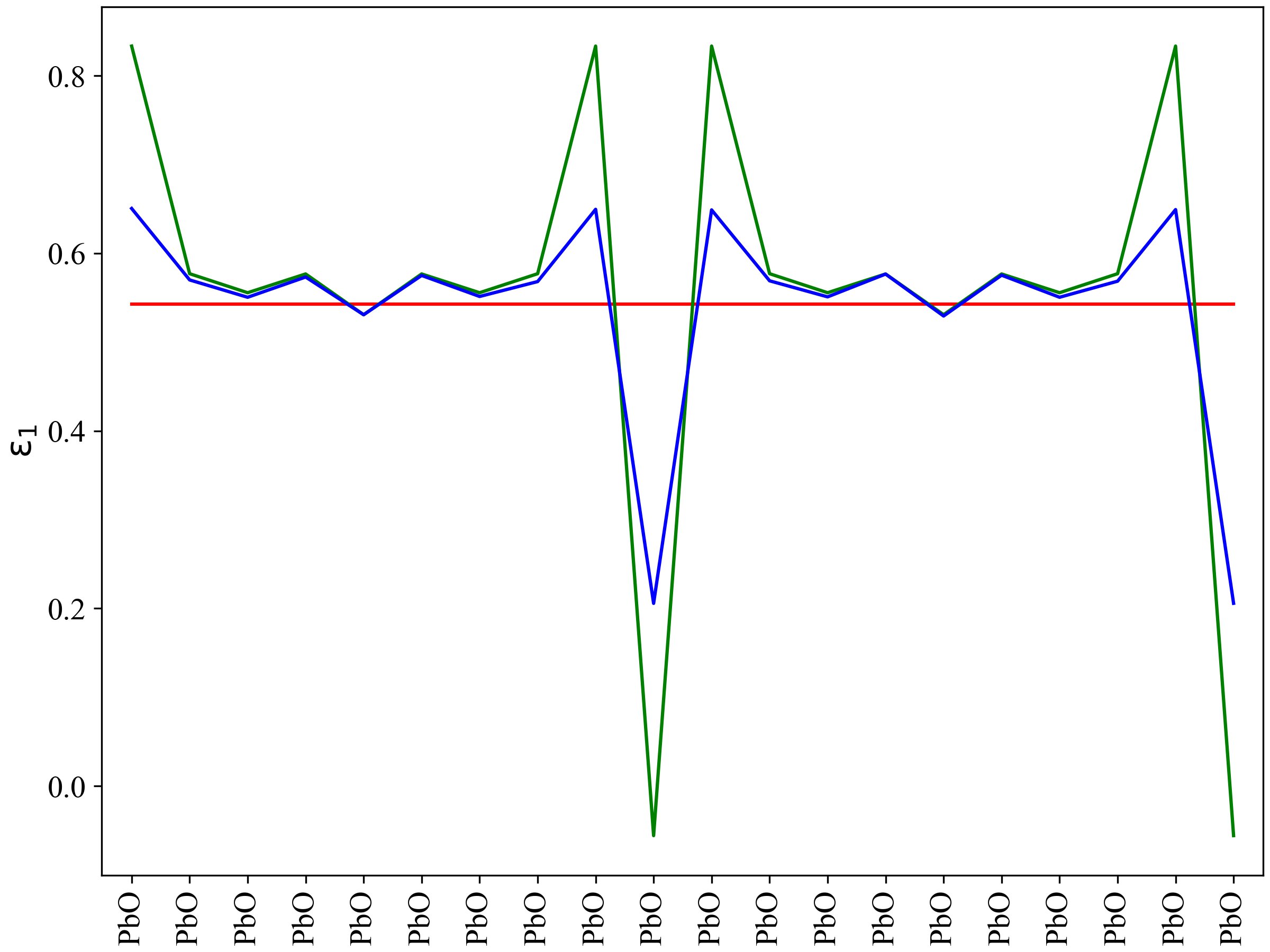}
    \caption{\justifying Layer-resolved profile of the in-plane strain component $\varepsilon_{1}$ (Voigt notation), evaluated with respect to the in-plane lattice constant of the bulk tetragonal ferroelectric phase. The strain is shown for three relaxation scenarios: (red) only the Ising component $P_{z}$ of the polarization is allowed to relax; (green) both the N\'eel ($P_{x}$) and Ising ($P_{z}$) components are relaxed; (blue) all three components of the polarization, including the Bloch component ($P_{y}$), are relaxed. In all cases, the lattice vectors are fully relaxed.
    }
 \label{fig:epsilon1}
\end{figure}

The presence of a N\'eel component, $P_x$, introduces position-dependent variations in the in-plane strain $\varepsilon_1$, as illustrated in Fig.~\ref{fig:epsilon1}. The associated strain gradient, $\eta_{xx,x}(x) = \partial \varepsilon_{xx}/\partial x$, becomes nonzero. While $P_x$ remains small (see inset of Fig.~\ref{fig:structure}), it peaks at the PbO planes where $\varepsilon_1$ reaches up to 0.8\%. Conversely, $\varepsilon_1$ drops to $-0.06\%$ at the domain wall, where $P_x$ vanishes. This inhomogeneous strain distribution helps relax the elastic energy and is reflected in the energetics summarized in Table-\ref{table:DW-energy} and discussed in Sec.~\ref{sec:dwenergy}. Importantly, the symmetry-adapted form of the flexoelectric tensor for the considered space group prohibits coupling between the strain gradient $\eta_{xx,x}$ and the Bloch component of the polarization, $P_y$.

\begin{figure}[H]
    \centering
    \includegraphics[width=\columnwidth]{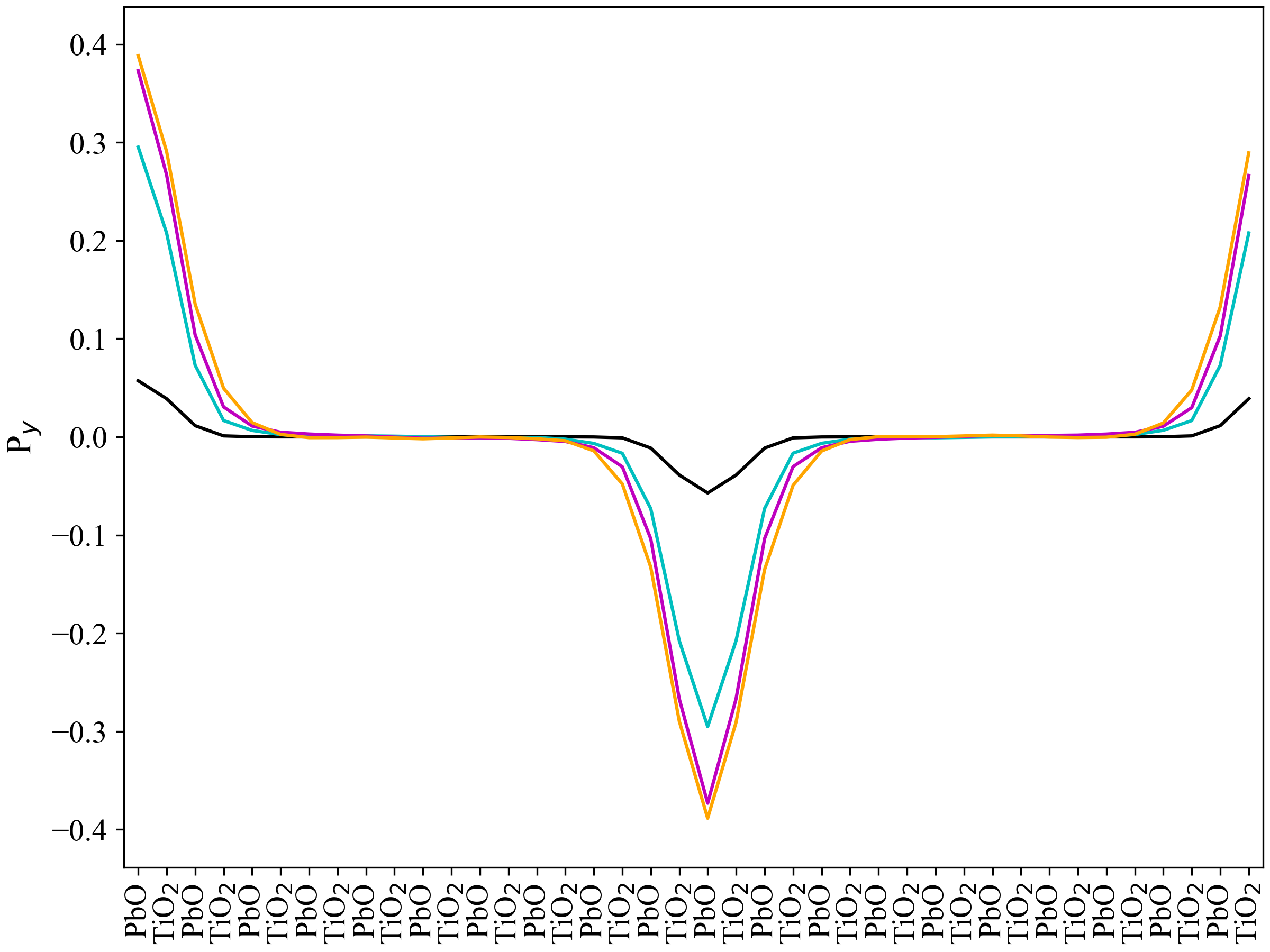}
    \caption{\justifying Layer-resolved profiles of the Bloch component of the polarization ($P_{y}$) for various locally stable domain wall structures, obtained from different initial configurations used in the structural relaxation. In all cases, the lattice vectors are fully relaxed. Polarizations are given in C/m$^{2}$.
    }
 \label{fig:localminimaBloch}
\end{figure}

Finally, under full relaxation, a Bloch component is clearly observed, whose magnitude increases with in-plane lattice constant. It originates from displacements of Pb atoms along the $y$ direction~\cite{Wojdel-14,Wang-14,Wang-17,Zatterin-24}. For the fully relaxed structure, the Pb displacement is $\Delta y_{\rm Pb} = 0.19$~\AA, decreasing to $\Delta y_{\rm Pb} = 0.18$~\AA\ under the SrTiO$_3$ epitaxial constraint, in agreement with prior {\sc abinit} results~\cite{Zatterin-24} (0.20~\AA\ and 0.19~\AA, respectively).
Interestingly, multiple local minima are found in {\sc siesta} depending on the initial Pb displacement relative to the N\'eel configuration (see Fig.~\ref{fig:localminimaBloch}). The energies reported in Table-\ref{table:DW-energy} correspond to the global minimum after systematic exploration of initial conditions. Despite substantial variation in the Bloch component, the Ising component, $P_z$, remains largely unchanged. Energy differences among metastable configurations are small -- of the order of 6~meV per 100-atom supercell -- indicating a flat energy landscape with respect to the development of the Bloch component.
A global minimum with consistent structural and energetic properties, as reported in Table-\ref{table:DW-energy}, can be reliably obtained by refining the real-space integration grid in {\sc siesta} and tightening relaxation thresholds.

The amplitude of the Bloch component at the minimum is also strongly dependent on the epitaxial strain, ranging from 0.35 C/m$^{2}$ when PbTiO$_{3}$ is strained on cubic SrTiO$_{3}$ to 0.42 C/m$^{2}$ when a DyScO$_{3}$ substrate is assumed. 

\subsection{Domain-wall energies}
\label{sec:dwenergy}

The domain wall energies for the supercells under different strain conditions are summarized in Table~\ref{table:DW-energy}.


\begin{table*}[!h]
   \begin{center}
     \caption{\justifying 
     Calculated domain wall energies using the PBEsol functional under various epitaxial strain conditions for a supercell of 20$\times$1$\times$1 u.c. (domain width of $10$ u.c.). Lattice constants $a$, $b$, and $c$ (in \AA) correspond to the three Cartesian directions.
     $P_{0}$ (in C/m$^{2}$) denotes the polarization at the center of the domain, and $\delta$ (in \AA) is the half-width of the domain wall, both obtained by fitting the polarization profiles shown in Fig.~\ref{fig:structure} to the functional form given in Eq.~(\ref{eq:funcform}). Domain wall energies are reported both in mJ/m$^{2}$, and meV/$\square$ (in brackets).
             }
     \label{table:DW-energy}
        \begin{tabular}{ >
        {\centering\arraybackslash}m{0.100\linewidth}  >
        {\centering\arraybackslash}m{0.100\linewidth}  >
        {\centering\arraybackslash}m{0.100\linewidth}  >
        {\centering\arraybackslash}m{0.100\linewidth}  >
        {\centering\arraybackslash}m{0.100\linewidth}  >
        {\centering\arraybackslash}m{0.145\linewidth}  >
        {\centering\arraybackslash}m{0.145\linewidth}  >
        {\centering\arraybackslash}m{0.145\linewidth} }
           \hline
           \hline
           $a$                                &
           $b$                                &
           $c$                                &
           $P_{0}$                            &
           $\delta$                           &
           $E_{\rm DW}^{\rm Ising}$           &
           $E_{\rm DW}^{\rm Ising}$ - $E_{\rm DW}^{\rm N\acute{e}el}$    &
           $E_{\rm DW}^{\rm N\acute{e}el}$ - $E_{\rm DW}^{\rm Bloch}$           \\
           \hline
           \multicolumn{8}{c}
           {Fully-relaxed configuration}      \\
           3.892 &
           3.893 &
           4.115 &
           0.99  &           
           2.24  &
           195.6  (195.4)                            &
           5.1    (5.1)                            &
           3.4    (3.3)                            \\
           3.884 &
           3.884 &
           4.122 &
           1.01  &           
           2.20  &
                                      &
                                       &
                                       \\
           3.892~\cite{Zatterin-24}         &
           3.897~\cite{Zatterin-24}         &
           4.075~\cite{Zatterin-24}         &
           &
           &
           179.5 (178.0)~\cite{Zatterin-24}         &
                                              &
           6.4 (6.5)~\cite{Zatterin-24}           \\
           &
           &
           &
           &
           &
           152 (148.0)~\cite{Wojdel-14}             &
                                              &
           4.1 (4.0)~\cite{Wojdel-14}             \\
           \hline
           \multicolumn{8}{c}
           {PbTiO$_{3}$ fixed to the optimized bulk tetragonal ferroelectric}    \\
           3.871 &
           3.871 &
           4.220 &
           1.15  &
           2.06  &
           223.1 (227.5)          &
           6.4  (6.6)           &
           2.2   (2.2)        \\
           \hline
           \multicolumn{8}{c}
           {PbTiO$_{3}$ strained on cubic SrTiO$_{3}$} \\
           3.889 &
           3.889 &
           4.120 &
           0.99  &
           2.21  &
           195.6  (195.6)          &
           5.0   (5.0)           &
           3.0    (2.9)          \\
           \hline
           \multicolumn{8}{c}
           {PbTiO$_{3}$ strained on cubic DyScO$_{3}$} \\
           3.912 &
           3.912 &
           4.071 &
           0.92  &
           2.39  &
           224.9  (223.5)         &
           5.0    (5.0)          &
           10.2   (10.0)          \\
           \hline
           \multicolumn{8}{c}
           {PbTiO$_{3}$ strained on orthorhombic DyScO$_{3}$}                      \\
           3.912 &
           3.916 &
           4.069 &
           0.92  &
           2.38  &
           230.4  (229.0)          &
           5.7    (5.6)          &
           10.7   (10.6)          \\
                      \hline
           \multicolumn{8}{c}
           {PbTiO$_{3}$ strained on orthorhombic DyScO$_{3}$}                      \\
           3.916 &
           3.912 &
           4.065 &
           0.91  &
           2.45  &
           229.4  (227.8)          &
           5.0    (5.0)          &
           9.8    (9.9)             \\
           \hline
           \hline
        \end{tabular}
   \end{center}
\end{table*}

It is important to note that domain walls (DWs) are associated with strain relaxations, which are inherently long-range in nature. 
As such, DW energies strongly evolves with the domain width~\cite{Poykko-99} and isolated DW energies are very hard to converge. Moreover, in systems with finite domain sizes, the concept of an isolated DW energy may not be the most relevant. In the present work, we consider stripe domain configurations with a domain width $t$ set by the supercell size $L$ (i.e., $t = L/2$).
\begin{figure}[H]
    \centering
    \includegraphics[width=\columnwidth]{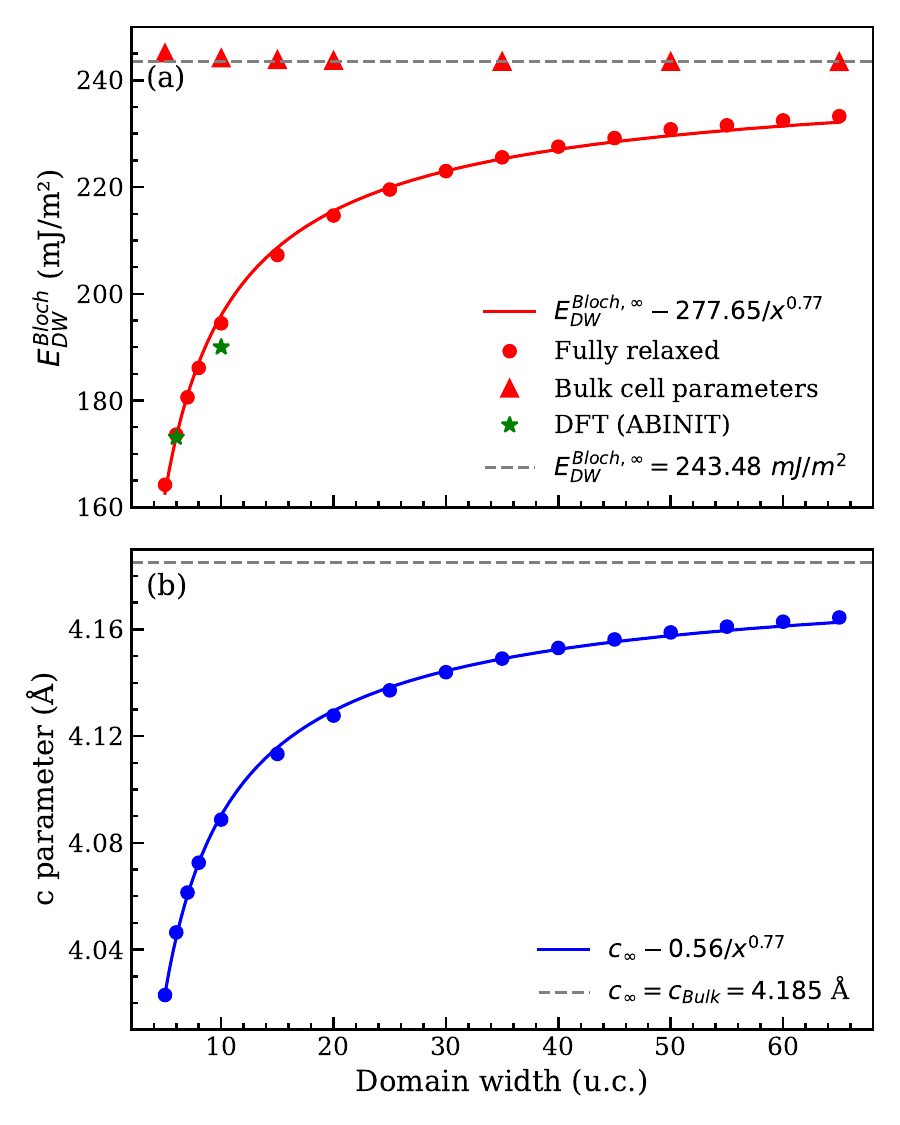}
    \caption{\justifying (a) Convergence of the Bloch DW energies with respect to the domain width $t$ in the fully relaxed case (red circles), constrained to the bulk tetragonal cell parameters case (red triangles) and \textsc{abinit} DFT fully relaxed results to which the model was fitted (green stars). The red curve represents a power-law fit to the energies on the fully relaxed case. (b) Convergence of $c$ lattice parameter with respect to the domain width (blue circles) fitted with a power-law (blue curve). 
    }
 \label{fig:conv_DW}
\end{figure}

In fully relaxed calculations—or under epitaxial constraints with out-of-plane relaxation of the lattice parameter $c$—increasing $t$ modifies the lattice parameters, which in turn influences the DW energy, $E_{\rm DW}$.

Alternatively, fixing the lattice parameters to their bulk values, as expected in the limit $t \rightarrow \infty$, provides a practical approach to estimate isolated DW energies from small supercells~\cite{MeyerVanderbilt2002}. Figure~\ref{fig:conv_DW} displays the evolution of $c$ and $E_{\rm DW}$ as a function of $t$, obtained from second-principles simulations employing the model of Ref.~\cite{Bastogne-25}. The results confirm that both quantities converge slowly when the lattice parameters are relaxed, exhibiting a power-law dependence on $t$. Consistent with previous findings~\cite{MeyerVanderbilt2002}, we also show that fixing the lattice constants to their bulk values yields DW energies in excellent agreement with the extrapolated isolated DW limit, even for relatively small supercells. Unless otherwise stated, all DW energies reported in this work correspond to $t = 10$ unit cells and are thus specific to that domain width.

Given these considerations, we can now examine the results. Within the PBEsol functional, the energy differences associated with the domain wall in the fully relaxed configuration are relatively large.
For the high-symmetry case, where only the Ising-type configuration is allowed in the fully relaxed PbO-centered domain wall, these energies range from 152 mJ/m$^{2}$ (148 meV/$\square$), as reported in Ref.~\cite{Wojdel-14}, to 195.6 mJ/m$^{2}$ (195.4 meV/$\square$) in our {\sc siesta} calculations for fully relaxed lattice constants in a $20 \times 1 \times 1$ supercell. Meyer and Vanderbilt obtained a lower value of 132 mJ/m$^{2}$ (128.5 meV/$\square$) at the LDA level using a $12 \times 1 \times 1$ and the lattice vectors constrained to the bulk tetragonal ferroelectric unit cell~\cite{MeyerVanderbilt2002}. These significant discrepancies are unexpected, given the overall good agreement in the polarization profile and domain wall polarization.
This variation can be attributed to differences in the DFT functional (LDA in Ref.~\cite{MeyerVanderbilt2002} and PBESOL in Refs.~\cite{Wojdel-14, Zatterin-24} and in this work), pseudopotentials (projector augmented wave method in Ref.~\cite{Wojdel-14} vs. optimized norm-conserving pseudopotentials used in this work and in Ref.~\cite{Zatterin-24}) and basis sets (plane waves in Refs.~\cite{Wojdel-14,Zatterin-24} vs. numerical atomic orbitals in this study). Notably, when the same pseudopotentials are used, the agreement in energy differences improves.

When a Néel-type polarization component is allowed at the domain wall, the energy decreases by approximately 5 mJ/m$^{2}$ (5.0 meV/$\square$), a value that is pretty constant independent of the strain conditions. 
This is surprising given that the Néel polarization at the domain wall is relatively small.
As discussed in Sec.~\ref{sec:dwstructure}, this is attributed to a relaxation of the elastic energy produced by an inhomogeneous strain in the plane.

Furthermore, the full development of polarization at the domain wall, including a Bloch component, leads to an additional energy reduction that strongly depends on the imposed in-plane lattice constant. In the fully relaxed structure, this reduction (with respect to the most relaxed N\'eel structure) amounts to 3.4 mJ/m$^{2}$ (3.3 meV/$\square$), a value that remains nearly unchanged when PbTiO$_{3}$ is strained atop a cubic SrTiO$_{3}$ substrate (with a lattice parameter of 3.899 \AA, as computed using the PBEsol functional~\cite{Zatterin-24}).
However, under a tensile strain similar to that induced by a DyScO$_{3}$ substrate, the energy reduction due to the Bloch component nearly triples with respect the value where only N\'eel polarization was allowed, increasing from 3.4 mJ/m$^{2}$ (3.3 meV/$\square$) to approximately 10.7 mJ/m$^{2}$ (10.6 meV/$\square$). 
This is somehow expected from previous studies on the phase-diagram of bulk PbTiO$_{3}$ under strain. 
Tensile strains favor the polarization rotation from a purely tetragonal $c$-phase to a phase where a coexistence of $c$-domains and $aa$ domains (with non-vanishing in-plane components) are favoured~\cite{Bungaro-04,Dieguez-05}. The fact that the P$_x$ values are not increasing with strain arises from symmetry constraints that enforce head-to-head and tail-to-tail polarization configurations, which are energetically unfavorable due to electrostatics.

This increase in the well-depth also suggests that the critical transition temperature below which a Bloch-like domain wall stabilizes is highly strain-dependent.
Second-principles simulations in Ref.~\cite{Zatterin-24}, suggested that $P_{y}$ became zero at approximately 150 K when PbTiO$_{3}$ was strained on cubic SrTiO$_{3}$, while for freestanding PbTiO$_{3}$ (no imposed strain that resulted in slightly larger in-plane lattice constants), a transition temperature equal to approximately 200 K was found. 
Further calculations with larger tensile strains should be performed for a proper numerical estimation of this temperature-strain dependence on the onset of the Bloch component.

\section{Conclusions}
\label{sec:conclusions}

In this work, we have investigated the structural and energetic properties of Bloch-type polarization components in 180$^\circ$ domain walls in bulk PbTiO$_{3}$ under varying mechanical boundary conditions, using first-principles simulations. Our results provide a comprehensive atomistic understanding of how epitaxial strain influences the polarization profile and energetics of domain walls in this prototypical ferroelectric.

We confirm that a Bloch component, primarily originating from displacements of Pb atoms along the domain wall plane, can spontaneously develop at $T$ = 0 K in otherwise Ising-type domain walls. The magnitude of this component, and the corresponding energy gain associated with its condensation, are found to be highly sensitive to the in-plane lattice constants. Specifically, tensile strains significantly enhance both the amplitude of the Bloch component and the associated energetic stabilization, with energy reductions reaching up to 10.7 mJ/m$^{2}$ (10.6 meV/$\square$) under strain conditions mimicking DyScO$_{3}$ substrates.

Our analysis highlights the delicate interplay between structural relaxations, strain-induced elastic energy variations, and polarization orientation at domain walls. While the Ising component remains robust, the Bloch component arises in a relatively flat energy landscape. This behavior suggests that external control via mechanical boundary conditions could serve as a viable route to engineer chiral ferroelectric textures in bulk PbTiO$_{3}$.

Finally, the strain dependence of the energy gain associated with the Bloch component provides a predictive framework to estimate the critical temperature for its stabilization, a critical magnitude to tailor topological and chiral functionalities in ferroelectric polar systems. 

%
%
%

\section{Acknowledgements}

S.C.~acknowledges financial support from Erasmus+ KA-107 action and the Vice-rectorate for Internationalisation and Global Engagement of the University of Cantabria.
J.J.~acknowledges financial support from Grant No.~PID2022-139776NB-C63 funded by MCIN/AEI/10.13039/501100011033 and by ERDF ``A way of making Europe'' by the European Union.
F.G.-O., L.B and Ph. G. acknowledge support by the European Union’s Horizon 2020 research and innovation program under Grant Agreement No. 964931 (TSAR).
F.G.O. also acknowledges financial support from MSCA-PF 101148906 funded by the European Union and the Fonds de la Recherche Scientifique (FNRS) through the grant FNRS-CR 1.B.227.25F. Ph. G. also acknowledges support from the Fonds de la Recherche Scientifique (FNRS) through the PDR project PROMOSPAN (Grant No. T.0107.20).
The authors also gratefully acknowledge the computer resources, technical expertise, and assistance provided by the Centre for High Performance Computing (CHPC-MATS862), Cape Town, South Africa. 

\newpage
\bibliography{bibliography}

\begin{thebibliography}{70}
\bibitem [{foo()}]{footnotepseudo}%
  \BibitemOpen
  \href@noop {} {}\bibinfo {note} {The scalar relativistic \textsc{oncvpsp}
  v0.4.1 pseudopotentials with stringent accuracy were used.}\BibitemShut
  {Stop}%
\end{thebibliography}

\end{document}